\documentclass[useAMS,usenatbib]{mn2e}

\usepackage{txfonts}
\usepackage[dvips]{graphicx}
\usepackage{epsfig}
\usepackage{float}

\usepackage{mathrsfs}

\def\lsim{\mathrel{\hbox{\rlap{\hbox{\lower4pt\hbox{$\sim$}}}\hbox{$<$}}}}
\def\gsim{\mathrel{\hbox{\rlap{\hbox{\lower4pt\hbox{$\sim$}}}\hbox{$>$}}}}
\def\kms {\rm{km~s^{-1}}}

\newcommand{\dunit}  {h$^{-1}\,$Mpc}

\title[Density and velocity void profiles in the SDSS]%
{Clues on void evolution II: Measuring density and velocity profiles
on SDSS galaxy redshift space distortions}

\author[Paz et al.]{
\parbox[t]{\textwidth}
{ 
  Dante Paz$^{1,2}$ \thanks{E-mail: dpaz@oac.uncor.edu},
  Marcelo Lares$^{1,2}$,
  Laura Ceccarelli$^{1,2}$,
  Nelson Padilla$^{3}$
  \& Diego Garc\'{\i}a Lambas$^{1,2}$
}
\vspace*{6pt}\\
$^1$ Instituto de Astronom\'\i a Te\'orica y Experimental, 
     UNC-CONICET, C\'ordoba, Argentina. \\
$^2$ Observatorio Astron\'omico de C\'ordoba, Universidad Nacional de C\'ordoba, Argentina. \\
$^3$ Departamento de Astronom\'\i a y Astrof\'{\i}sica, Pontificia
     Universidad Cat\'olica de Chile, Santiago, Chile.\\
}

\begin{document}

\date{\today}

\maketitle

\begin{abstract}
%{{{*/
%
Using the redshift-space distortions of void-galaxy cross-correlation function
we analyse the dynamics of voids embedded in different environments.
We compute the void-galaxy cross-correlation function in the Sloan Digital Sky
Survey (SDSS) in terms of distances taken along the line of sight and projected
into the sky.
We analyse the distortions on the cross-correlation isodensity levels and we
find anisotropic isocontours consistent with expansion for large voids with
smoothly rising density profiles and collapse for small voids with overdense
shells surrounding them.
Based on the linear approach of gravitational collapse theory we developed a
parametric model of the void-galaxy redshift space cross-correlation function.
We show that this model can be used to successfully recover the underlying
velocity and density profiles of voids from redshift space samples.
By applying this technique to real data, we confirm the twofold nature of void
dynamics: large voids typically are in an expansion phase whereas small voids
tend to be surrounded by overdense and collapsing regions.
These results are obtained from the SDSS spectroscopic galaxy catalogue and
also from semi-analytic mock galaxy catalogues, thus supporting the viability
of the standard $\Lambda$CDM model to reproduce large scale structure and
dynamics.
%
%}}}*/
\end{abstract} 


\section{Introduction} \label{S_intro}
%{{{*/

Large scale underdensities naturally arise as the result of structure
growth.
According to our current understanding, the structure of matter evolves 
from small density fluctuations in the early universe to build up the
present day distribution of matter.
As the universe evolves, galaxies dissipate from underdense regions
and progress towards matter concentrations by the action of gravity,
forming both voids and filaments in the process.
The void distribution evolves as matter collapses to the structure and
galaxies dissipate from voids, making a supercluster-void network
\citep{frisch_evolution_1995,einasto_supercluster-void_1997,einasto_multimodality_2012}.  
This interplay in the formation of voids and structures, allows to
think of them as complementary, both encoding useful information to place 
constraints on the parameters of cosmological models. 
Hence, while the predominant objects of the large scale galaxy distribution
are structures such as groups, clusters, filaments or walls, voids
emerge as the relevant features that shape, along with filaments, the 
structure at the largest scales.

Underdense regions have been identified and analyzed in numerical 
simulations  \citep{
   hoffman_origin_1982,
   hausman_evolution_1983,
   fillmore_self-similar_1984,
   icke_voids_1984,
   bertschinger_self-similar_1985,
 aragon-calvo_unfolding_2010,
 aragon-calvo_hierarchical_2013,
kauffmann_voids_1991}
   and in galaxy catalogues 
   \citep{
 pellegrini_voids_1989,
 slezak_objective_1993,
   el-ad_voids_1997,
   el-ad_catalogue_1997, 
   el-ad_case_2000, 
 muller_voids_2000,
 plionis_size_2002,
 hoyle_voids_2002,
   hoyle_voids_2004,
   ceccarelli_voids_2006, 
 patiri_statistics_2006,
 neyrinck_zobov:_2008}
showing similar properties regardless of the details of the
identification methods \citep{colberg_aspen-amsterdam_2008} and galaxy 
sample properties.

\citet{padilla_spatial_2005} show that voids defined by the spatial
distribution of haloes and galaxies have similar statistical and
dynamical properties.
Moreover, the statistics of void and matter distributions are strongly
related \citep{white_hierarchy_1979} and therefore voids are a
powerful tool to study the formation and evolution of overdense
structures. 
Since the void population properties are sensitive to the details of
structure formation, they can be used to constrain cosmological models
\citep[e.g.][]{peebles_void_2001, park_challenge_2012,
   kolokotronis_supercluster_2002, colberg_voids_2005,
lavaux_precision_2010, bos_darkness_2012, biswas_voids_2010,
benson_galaxy_2003, park_challenge_2012, bos_less_2012,
hernandez-monteagudo_signature_2012, clampitt_voids_2013}.
Also, due to the global low density environment in which void galaxies are 
embedded, the galaxy populations in or close to voids are valuable to
shed light on the mechanisms of galaxy evolution and its dependence on
the large scale environment \citep{lietzen_environments_2012,
hahn_properties_2007, hahn_evolution_2007, ceccarelli_low_2012,
ceccarelli_large-scale_2008, gonzalez_galaxy_2009}.

The simplest approach allows us to characterize individual voids as
spherical regions with isotropic motions \citep{icke_voids_1984,
van_de_weygaert_peak_1996, padilla_spatial_2005,
ceccarelli_voids_2006}.
However, more detailed analyses suggest that voids are not isolated
structures but form part of an intricate network which affects their
dynamical properties \citep{ bertschinger_self-similar_1985,
   melott_generation_1990, mathis_voids_2002,
colberg_intercluster_2005, shandarin_shapes_2006,
platen_alignment_2008, aragon-calvo_hierarchical_2013,
patiri_quantifying_2012}.

In a previous work \citep[][hereafter Paper I]{ceccarelli_clues_2013},
we performed a statistical study of the void phenomenon focussing on
void environments.
To that end, we examined the distribution of galaxies around voids in
the SDSS by computing their integrated density contrast profile. 
By defining a separation criterion to characterize voids according to
their surrounding environment, we obtained two characteristic void 
types, according to their large-scale radial density profiles:
(i) Voids with a density profile indicating an underdense region 
surrounded by an overdense shell, were dubbed S-Type voids; 
(ii) voids showing a continuously rising density
profiles were defined as R-Type voids.  
We also found that small voids are more frequently surrounded by
overdense shells, and thus they are typically S-type. 
On the other hand, larger voids are more likely classified as R-Types, 
i.e., with an increasing integrated density contrast profile, which 
smoothly rises towards the mean galaxy density.
Moreover, this behaviour of SDSS voids results in a correlation 
between the fraction of voids surrounded by overdense shells and
their sizes. This fraction continuously decreases as the void size
increases, in a similar way for real, mock and direct numerical simulation samples.

Such a dichotomy in the behaviour of voids was first introduced by 
\citet{sheth_hierarchy_2004}, based on an excursion set formalism.
The authors classify void profiles and relate them to one of two
processes: 
The void-in-void process describes the evolution of voids that are
embedded in larger-scale underdensities.
This is the case when small voids merge at an early epoch with other
void to form a larger void at a later epoch.
On the other hand, underdense regions embedded within larger overdense
regions undergo a so-called void-in-cloud process.
In a hierarchical structure formation scenario, the filament
network subtended by dark matter halos is modified by halo merging.
Eventually, some voids located in the interstices of this network
will shrink at later times constituting the void-in-cloud scenario. 
This last case seems to affect more likely small rather than large voids.

Since the evolution of structure in the universe shapes the large 
scale matter clumps and the voids at the same time, both types of
structures are responsive to the details of the contents of the
universe, and the equation of state of its constituent species
\citep{einasto_towards_2011}.
Furthermore, the physics of galaxies in voids is simpler since the
non-linear effects of gravity are less significant in regions of space
devoid of galaxies.
The evolution of void galaxies is affected by the surrounding
environment when galaxies are located close to the void edge
\citep{lindner_distribution_1996, ceccarelli_low_2012}.
Assuming spherical symmetry, \citet{fillmore_self-similar_1984} derive
similarity solutions to describe the evolution of voids in a perturbed
Einstein-de Sitter universe filled with cold, collisionless matter.
They suggest, for this simplified model, that different void 
modes would arise depending on the steepness of the initial density
deficit.
As a result, 
the statistics of the void population has been used to constrain
parameters of the standard cosmological model \citep{
betancort-rijo_statistics_2009, biswas_voids_2010, bos_darkness_2012,
bos_less_2012},
and void catalogues have been exploited to test alternative 
cosmological models \citep{ biswas_swiss-cheese_2008,
bolejko_formation_2005, clampitt_voids_2013}.

Studies on the dynamics (and evolution) of regions around cosmological
voids have been implemented mainly in numerical simulations and
semi-analytical galaxies by several authors. 
For instance, \citet{regos_evolution_1991} have studied the evolution
of voids in numerical simulations obtaining the peculiar streaming
velocities of void walls.
\citet{dubinski_void_1993} and \citet{padilla_spatial_2005} have
analysed the peculiar velocity field surrounding voids in simulations.
Also, \citet{sheth_hierarchy_2004} and
\citet{paranjape_hierarchy_2012} studied the void size evolution in
simulations.
\citet{aragon-calvo_hierarchical_2013} examined the internal dynamics
of voids and their hierarchical features.
Albeit, the dynamics of voids have not been extensively studied on
observational data.
\citet{ceccarelli_voids_2006} used redshift space distortions,
peculiar velocities, and a non-linear approximation to determine
properties of the peculiar velocity field around voids in the 2dfGRS,
including the amplitude of the expansion of voids and the dispersion
of galaxies in the directions parallel and perpendicular to the void
walls.
\citet{patiri_quantifying_2012} suggest the presence of coherent
outflows of galaxies in the vicinity of large voids in SDSS.

This paper is organized as follows.
In Section \ref{S_data} we describe the galaxy samples
and the corresponding void catalogues. 
We also describe the semi-analytic mock galaxy samples built from the 
simulation box.
The redshift space distortions on the correlation function are 
analyzed in Section \ref{S_xi_data}.
In Section \ref{S_model} we present the theoretical approach
adopted to model the redshift space distortions from density profiles and velocity flows 
of galaxies around voids. 
A comparison of observational results to the numerical simulation and
the mock catalogue is given in Section \ref{S_test}, and the results
obtained for the observational data are shown in Section \ref{S_results}.
Finally, we discuss our results in Section \ref{S_concl}.  
%}}}*/
 
% FIG 1
%{{{*/ 
%_____________________________________________________________FIG.1
\begin{figure*}
\includegraphics[width=0.75\textwidth]{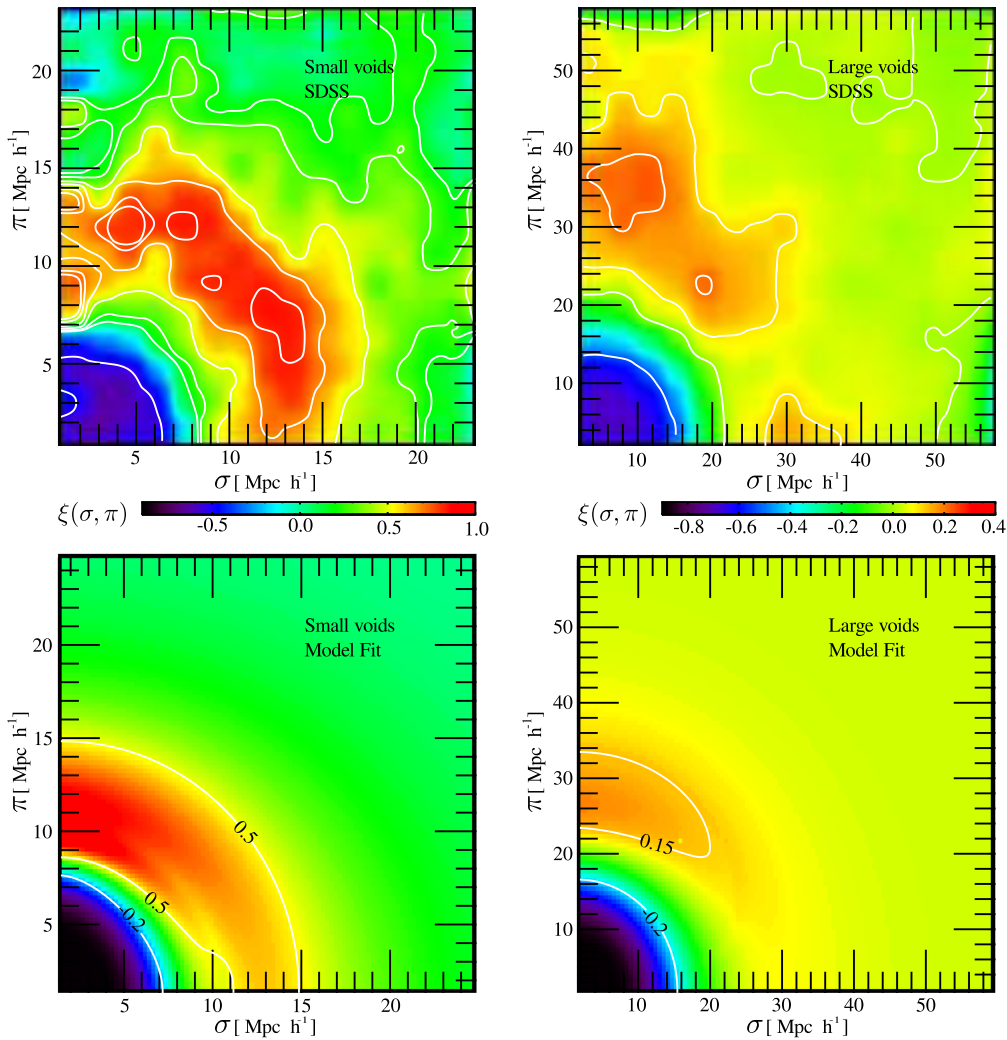}%
\caption{
   Redshift space distortions of galaxies in the SDSS (upper panels) and
   the best fit models (bottom panels) for small S-type voids in S1 
   ($6<\mathrm{R}_{\mathrm{void}}/$\dunit$<8$, left panels)
   and large R-type voids in S3
   ($10<\mathrm{R}_{\mathrm{void}}/$\dunit$<20$, right panels).
   The fractions of small R-type 
   voids and large S-type voids are 0.3 and 0.1, respectively. 
   Notice that spatial and color scales change for small and large
   voids.
}
\label{F_SDSS_xisigpi} 
\end{figure*}

%________________________________________________________________.     
%}}}*/

\section{Data sets} \label{S_data}
%{{{*/

We use the Main Galaxy Sample \citep{strauss_spectroscopic_2002} from
the Sloan Digital Sky Survey data release 7
\citep[\textsc{SDSS-DR7,}][]{abazajian_seventh_2009}.
SDSS photometric data provides CCD imaging data in five photometric
bands \citep[UGRIZ,][]{fukugita_sloan_1996, smith_ugriz_2002}.
The \textsc{SDSS-DR7} spectroscopic catalogue comprises in this
release 929,555 galaxies with a limiting magnitude of
\mbox{$r~\leq~17.77$ mag}.

We perform the identification of voids using the algorithm presented
by \citet{padilla_spatial_2005} and tested in
\citet{ceccarelli_voids_2006}.
The general properties of the SDSS void sample are introduced in Paper
I.
Voids in the galaxy distributions are identified over three different
volume complete samples, with limiting redshift $0.08$, $0.12$ and
$0.15$, whereas corresponding maximum absolute magnitudes in the
$r-band$ are $M_r=-19.2$, $-20.3$ and $-20.8$, respectively.
We denote these three samples as S1, S2 and S3 (see Table \ref{T_SampleDef}). 
In order to compute absolute magnitudes needed in sample definitions,
we use the same cosmological parameters than that of the simulation,
described later on this Section.
The algorithm starts with the identification of the largest spherical 
regions where the overall density contrast is at most $\delta=-0.9$.
The list is cleaned so that each resulting spherical region is not
contained in any other sphere satisfying the same condition. 
The method also avoids the selection of spheres closer than two maximum 
void radii from the survey boundaries.
The centres of underdense spheres are chosen as the locations of void
centers, and the scale assigned to each void is the radius of the
underdense sphere.
It should be noticed that this procedure does not assume that voids
are spherical, but ensures that the void is surrounded by a spherical
region with overall density below a threshold of $0.1$ times the mean 
density.
Since the resulting void sample depends on the sample dilution
\citep{padilla_spatial_2005}, we seek for the best compromise between
the void sample size and the identification confidence, specially for
the smallest voids.
Therefore, the limiting redshift of the sample is chosen so that a
good quality of the void sample is obtained, and also the number of
voids remains large enough to achieve statistically significant
results.
With these criteria, voids down to \mbox{5 \dunit} are well resolved in
S1 sample of the catalogue.
We obtain 131 voids in this sample with radii ranging from 
\mbox{5 \dunit} to \mbox{22 \dunit}.
The smallest voids are identified in S1, since it has the 
greater galaxy density.
However, due to the limited volume, this sample is not suitable to
perform statistical studies of the largest voids, and for this we turn to the
additional samples, S2 and S3.
The number of voids and the definition of each sample are indicated in
Table \ref{T_SampleDef}.
It can be noticed that the intermediate redshift sample contains a mix 
of both small and large voids.

In order to test the results, we will implement our method on void
samples extracted from a mock galaxy catalogue and from a simulation
box.
We use galaxies from the semi-analytic model of galaxy formation by
\citet{bower_flip_2008} run on top of the Millennium simulation
\citep{springel_simulations_2005, lemson_halo_2006}.
The Millennium cosmological simulation adopts a $\Lambda$CDM
cosmological model and follows the evolution of $2160^3$ particles,
each with 8.6 $\times\, 10^8\, \mathrm h^{-1}M_{\odot}$ in a comoving
box of \mbox{500 $\mathrm{Mpc}$} a side. 
The parameters of the model, based on WMAP observations
\citep{spergel_first-year_2003} and the 2dF Galaxy Redshift Survey
\citep{colless_2df_2001}, are \mbox{$\Omega_{\Lambda}$ = 0.75},
\mbox{$\Omega_M$ = 0.25}, \mbox{$\Omega_b$ = 0.045}, \mbox{h = 0.73},
\mbox{n = 1} and \mbox{$\sigma_8$ = 0.9}. 
The semi-analytic model of galaxy formation
\citep[\textsc{GALFORM},][]{bower_flip_2008}, generates a population
of galaxies within the simulation box, by following the simulated
growth of galaxies within dark matter haloes in the simulation.
We identified voids in the full simulation box, taking into account
the fact that the minimum size of voids that the algorithm is capable
of identifing depends on the mean galaxy density
\citep{padilla_spatial_2005}.
Accordingly, we imposed a magnitude cut in the sample of semi-analytic galaxies 
diluting it so that the
mean galaxy density is the same than that of the SDSS sample.
The semi-analytic galaxy catalogue is found to contain 2534 voids with
sizes ranging from 5 to \mbox{22 \dunit} (see Table
\ref{T_SampleDef}).

The mock catalogue is constructed by reproducing the selection
function and angular mask of the SDSS.
The resulting semi-analytic galaxy catalogue has similar properties 
and observational biases to those of the SDSS catalogue.
We will use this catalogue in order to calibrate our statistical
methods, to interpret the data, and to detect any systematic biases in
our procedure. 
Mainly, we use positions in real space and peculiar velocities of
galaxies to test possible projection biases and to quantify the
effects of redshift space distortions.
The semi-analytic galaxy dataset also provides information on SDSS photometric 
magnitudes, star formation rates and total stellar masses, based on
computations from the semi--analytic model of galaxy formation
\citep{bower_flip_2008}.
The details of the number of R and S-type voids are given in Table
\ref{T_SampleDef}.
Following the same procedure carried out in the SDSS, voids 
are identified on three volume limited samples, with the same redshift
and magnitude thresholds than the used on real data.  (i.e. maximum
redshift of $0.08$, $0.12$ and $0.15$ and maximum $r-band$ magnitude
of $-19.2$, $-20.3$ and $-20.8$, respectively).
We denote these three samples as M1, M2 and M3, 
and comprise 113, 232 and 316 
voids, respectively. 

From pondering the void radii distributions obtained in each sample
we conclude that different samples are more suitable to study voids of
different sizes.
In the samples with the smallest volumes (S1 and M1) we do not find a
significant number of voids with radii larger than \mbox{12 \dunit}.
However, a significant number of voids with radii smaller than about
\mbox{10 \dunit} are identified, making those samples
more suitable for analysing
voids of radii between 5 and \mbox{10 \dunit} rather than larger voids.
The intermediate volume samples (S2 and M2) reach the maximum
number of voids with radii in the range \mbox{12--15 \dunit}.
These
samples are the most appropriate for studying voids of 
intermediate size
rather than either smaller or larger voids.
In the most extensive samples (S3 and M3) we find the largest number
of large voids \mbox{(R$_{void}$ $>$ 15 \dunit)} whereas 
the number of small voids is not adequate for statistical analyses.
According to this, we prefer the small volume samples for
a detailed statistical study of small voids while large volume samples
are used to examine the properties of large voids.

%}}}*/

% Table 1
%{{{*/
\begin{table}
\begin{minipage}{0.5\textwidth}
\begin{tabular*}{0.5\textwidth}{@{\extracolsep{\fill}}cccccc}
 \cline{1-6}
& \multicolumn{3}{c}{selection criteria} & &  \\
\cline{2-4}
sample & parent catalogue & limiting magnitude & z$_{lim}$ & N$_{S}$ & N$_{R}$ \\
\cline{1-6} \\

S1 & SDSS-DR7 & -19.2 & 0.08  & 48  & 83  \\
S2 & SDSS-DR7 & -20.3 & 0.10  & 73  & 174 \\
S3 & SDSS-DR7 & -20.8 & 0.12  & 71  & 252 \\
\\
M1 & mock & -19.2 & 0.08  & 48  & 65  \\
M2 & mock & -20.3 & 0.10  & 70  & 162 \\
M3 & mock & -20.8 & 0.12  & 109 & 207 \\
\\
   & simulation box & -15.9 & - & 1691 & 843 \\

\\ \cline{1-6}
\end{tabular*}
\caption{
Galaxy sample selection limits and their corresponding void 
samples.  In all cases we use in void identification volume limited samples, defined by a
maximum redshift ($z_{lim}$) and a limiting absolute magnitude in the
r-band.
The magnitude of galaxies in the mock and simulation box
samples correspond to that of the semi-analytic galaxy catalogue.
The number of R-type voids (N$_R$) and S-type voids (N$_S$) are also
indicated.
}
\label{T_SampleDef} 
\end{minipage}
\end{table}
%}}}*/

\section{SDSS void galaxy cross-correlation function}
\label{S_xi_data}
%{{{*/

In modern spectroscopic galaxy catalogues, redshift measurements are
commonly used to estimate galaxy distances.
However, these quantities include a contribution from the peculiar velocity 
component in the line of sight.
While this is a drawback when trying to obtain an accurate
three-dimensional map of the local universe, dynamical studies can
take advantage of the distortions imprinted in redshift space to
obtain information about the velocities.
In Paper I we have analyzed the large scale
environment around voids. 
We conclude that it is expected that the differences in
the spatial distribution of galaxies around voids also manifests as 
differences in the dynamical properties.
Consequently, it is natural to infer that redshift space distortions
on the correlation function will show these differences. 
In this section we search for the dynamics of voids in the redshift space
distribution of galaxies around them.

To this end, we measure 
the void-galaxy
cross-correlation function $\xi(\sigma,\pi)$ as a function of the
projected ($\sigma$) and line of sight ($\pi$) distances to the void
centre.
The $\xi(\sigma,\pi)$ function is the excess in the probability of
having a galaxy around a given void centre. 
The standard method to estimate such probability is by counting
void-galaxy pairs and normalising by the expected number of pairs for
a homogeneous distribution.
To compute this normalization it is necessary to produce a random
distribution of points with the survey selection function.
There are several estimators based on this counting procedure.
In this work we have evaluated two of them, the classic estimator
\citep{davis_survey_1983} $\xi=DD/DR-1$, where $DD$ and $DR$ are the
numbers of {\it void-galaxy} and {\it void-random tracer} pair counts
respectively, and a symmetric version of the \cite{landy_bias_1993}
estimator.
The latter is computed as $\xi=(DD-DR-RD+RR)/RR$, where in 
addition to $DD$ and $DR$, we need to calculate $RR$ and $RD$, the
numbers of {\it random centre-random tracer} and {\it random
centre-galaxy} pair counts respectively. 
Since we use volume limited samples of voids the random centre 
distribution is uniform. 
We found negligible differences between these two estimators for all
void samples. 
Thus for the sake of simplicity, we perform all the analysis using the
\citeauthor{davis_survey_1983} estimator.
The random sample of tracers needed for this estimator was generated
following the same procedure described in \citet{paz_alignments_2011}. 
Briefly, the expected numerical density of galaxies at a given
redshift, with a magnitude below the limit of the survey, is computed
from a Schechter luminosity distribution with parameters
$\phi_*=0.0149 \;, M_*=-20.44\;, \alpha=-1.05$
\citep{blanton_galaxy_2003}. 
The angular selection of the random points consists of a pixel mask 
based on the \textsc{SDSSPix} software \citep{swanson_methods_2008}.
This random catalogue contains about $2\times10^7$ random points
\citep[see][for more details]{paz_alignments_2011}.

Without redshift space distortions, $\xi(\sigma,\pi)$ would be
isotropic as a function of $\pi$ and $\sigma$. 
Therefore, any observed anisotropy in the measured $\xi$ function
would be evidence of the presence of line-of-sight velocities.
Given that these velocities only affect the $\pi$ scale, $\xi$ as a
function of $\sigma$, at low $\pi$ values, resembles the real space
correlation function.
On the other hand, the behaviour of $\xi$  as a function of $\pi$, for 
low $\sigma$ values, is fully affected by redshift space distortions.

We have defined two types of centre voids depending on their density profiles.
Hereby we briefly describe the procedure carried out to define the centre
samples; more details can be found in Paper I. 
The mean integrated density contrast profiles can be defined for
$\mathrm{R}_{\mathrm{void}}$ intervals.
As shown in Paper I, these average curves have a well defined maximum at a
distance $\mathrm{d}_{max}$ from the void centre, except for the largest voids
that exhibit an asymptotically increasing profile.
We classify voids into two subsamples according to positive or negative values
of the integrated density contrast at $\mathrm{d}_{max}$.
Voids surrounded by an overdense shell are dubbed S-Type voids, and satisfy 
$\Delta(\mathrm{d}_{max}) > 0$.
On the other hand R-Type voids are defined as those that satisfy the condition
$\Delta (\mathrm{d}_{max}) < 0$, which corresponds to voids with continuously rising
density profiles.
This scheme is applied for voids in each of the three volume limited subsamples
of the SDSS and mock catalogues, as defined in Section \ref{S_data}, as well as
in the semi-analytic sample of galaxies in the simulation box.

In the upper panels of Fig. \ref{F_SDSS_xisigpi} we show the void-galaxy
cross-correlation function of voids in the SDSS.
The upper left panel shows the correlation function for the sample of small S-type
voids in sample S1.
This is representative of small voids since the sample of voids with $\mathrm{R}_{\mathrm{void}}$
in the range 6--8 \dunit is dominated by S-type voids (80\%).
As can be seen in this panel, there is a clear excess in the number counts
of void-galaxy pairs at distances larger than $\mathrm{R}_{\mathrm{void}}=8$ \dunit (red colours), 
produced by the characteristic shell of these
samples of voids. 
It can be noticed that there is a compression of the isocorrelation curves in the $\pi$
direction.
This is an indication of the mean flow of galaxies towards the void centre.

In the upper right panel we show the correlation function for the
sample of large R-type voids in S3. 
Notice that the spatial and color scales are not the same in each panel to
make the variations between each case more clearly visible. 
As can be seen, the excess of void-galaxy pair counts in the 
case of small voids at scales \mbox{10--15 \dunit} is not present in the
sample of larger voids.
Given the trend in the fraction of S-type voids as a function of void
radius (reported in Paper I), the sample of large voids is dominated by R-type voids. 
A continuously rising profile is expected in this case, and indeed it
is observed in the $\sigma$ axis at low $\pi$ values.
However, as it can be seen in the upper right panel of this figure, an
asymmetric structure appears, which is clearly originated on redshift space distortions.

We also show, in the bottom panels of Fig. \ref{F_SDSS_xisigpi}, the synthetic
$\xi(\sigma,\pi)$ functions obtained after the application of a model to the
corresponding observed functions in the upper panels.  We present and describe
this model in the following section, where it is used to analyse in more detail
the dynamics obtained from redshift distortions.

%}}}*/

% FIG 2
%{{{*/ 
\begin{figure*}
\includegraphics[width=0.75\textwidth]{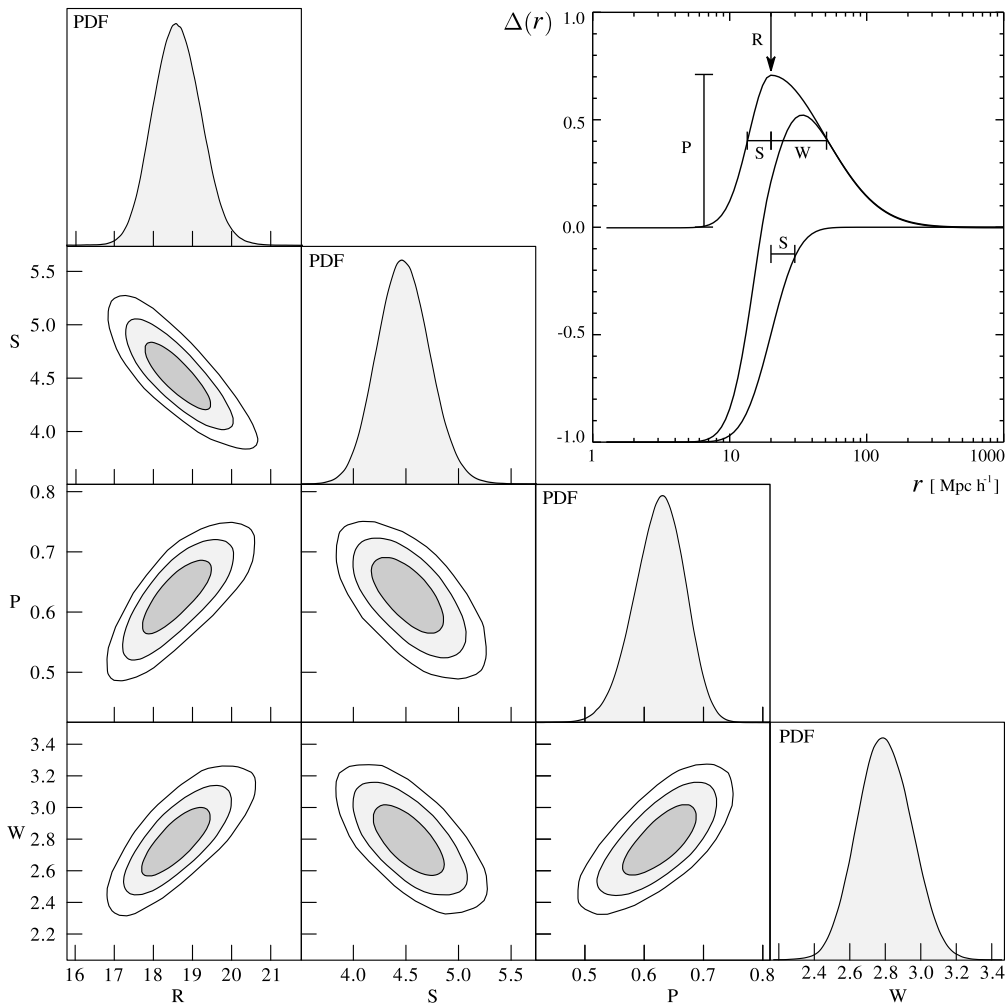}%
\caption{
   Projections of the Likelihood function for the model applied to
   intermediate size voids in the mock catalogue. 
   Voids have been identified in the M2 sample of the mock catalogue
   with radii in the range of \mbox{10--12 \dunit}.
   The upper-right panel schematically shows the model for the density
   profiles with a representation of the meaning of each of the
   parameters in our model.}
\label{lklyhd_12-10_12-over_mock}
\end{figure*} %}}}*/ 

\section{Model for $\xi(\sigma,\pi)$} \label{S_model}
%{{{*/

In the previous section we presented the correlation function obtained 
from two particularly interesting samples of voids. As has been shown, these samples
exhibit different distortion maps.
In
order to go deeper in the interpretation of such anisotropies, we have
implemented a model of the redshift space distortions on the void-galaxy
cross-correlation function. Following \citet{peebles_mean_1979} we compute the
$\xi(\sigma,\pi)$ function as the convolution of the real space correlation,
$\xi(r)$, and the pairwise velocity distribution, $g(\bmath{r},\bmath{w})$:
\begin{equation}
	\label{main_eq}
	1+\xi(\sigma,\pi)=\int{d^3w}\,g(\bmath{r},\bmath{w})\left[1+\xi(r)\right],
\end{equation}

\noindent
where $\bmath{r}\equiv(r_1,r_2,r_3)$ is the real space position of the tracer galaxy with
respect to the centre void and $\bmath{w}=\Delta\bmath{v}$ is the velocity of the tracer in the
rest frame of the centre object (pairwise velocity).  Subscripts denote each 
of the Cartesian coordinates: the third axis is taken along the line of sight
whereas $r_1$ and $r_2$ are coordinates in the plane of the sky.  The redshift
space separations of the void-galaxy pair are:

\begin{eqnarray}
	\sigma=\sqrt{r_1^2+r_2^2}\,\,\,\mathrm{and}\,\,\,\pi=r_3+w_3/\mathrm{H},\nonumber
\end{eqnarray}
which are parallel and perpendicular to plane of the sky, respectively (as
defined in the previous section).  As mentioned before, the third component of
the pairwise velocity ($w_3/\mathrm{H}$ in scale units, where H is the Hubble parameter at present time) is the source of the difference
between real and redshift space line of sight separations ($r_3$ and $\pi$).

In order to compute the redshift space correlation function it is necessary to adopt
some prescription for the pairwise velocity distribution, $g$. We assume that
this function can be approximated as a Maxwell-Boltzmann distribution centered
on a mean velocity field. 
The latter is a bulk flow given by linear theory, where the 
mean velocity of the distribution is a function of the density \citep{peebles_peculiar_1976}.
Since the velocities over the plane of
the sky, $w_1$ and $w_2$, do not affect $\sigma$ or $\pi$ coordinates, the
model only requires the definition of the marginal distribution $f$: 
\begin{eqnarray}
	f\left(w_3-\frac{r_3}{r}v(r)\right)&=&
	\iint dw_1\,dw_2\,g\left(\bmath{w} - \frac{\bmath{r}}{r}v(r)\right)\nonumber\\
	&=&\frac{1}{\sqrt{2\pi}{\sigma}_v}\mathrm{exp}\left(\frac{-\left(w_3 - v(r)\frac{r_3}{r}\right)^2}{2{\sigma}_v}\right).\nonumber
\end{eqnarray}
Finally, the correlation function in redshift space is obtained from 
\begin{equation}
	\label{xi_int}
	1+\xi(\sigma,\pi)=\int{dw_3}\frac{1}{\sqrt{2\pi}{\sigma}_v}\mathrm{exp}\left(\frac{-\left(w_3 - v(r)\frac{r_3}{r}\right)^2}{2{\sigma}_v}\right)\left[1+\xi(r)\right],
\end{equation}
where $r_3=\pi-w_3/\mathrm{H}$ and $r^2=\sigma^2+\left(\pi^2-\frac{w_3}{\mathrm{H}}\right)^2$.

We denote the mass density contrast within a sphere of radius $r$ as $\Delta(r)$. 
Following \citet{peebles_peculiar_1976}, the mean radial velocity $v(r)$ is related
to $\Delta(r)$ by the linear approximation:
\begin{equation}
	\label{v_eq}
	v(r)\approx-\mathrm{H}r\Delta(r)\frac{\Omega_m^{0.6}}{3}.
\end{equation}
We have tested other non-linear prescriptions relating $\Delta$ and $v$, 
given by \citet{yahil_dynamics_1985} and \citet{croft_apm_1999}.
However we have not found any significant difference
with the linear treatment.
This resides in the fact that the density contrast remains small up to large scales in areas around voids.

The integrated density contrast $\Delta(r)$ in a void centered sphere
of radius $r$ and volume $V$, is:

\begin{eqnarray}
      \Delta(r) &=& \frac{1}{V} \int_{V} \frac{\rho(r)}{\bar{\rho}}
      dV - 1
   \nonumber \\
      &=& \frac{3}{r^3} \int_0^r \xi(r) r^2 dr,
   \nonumber
\end{eqnarray}
where we have used that $\displaystyle \xi(r)=\rho/\bar{\rho} - 1$,
for a cross--correlation function $\xi(r)$.
Then, the real space void-galaxy cross correlation function is related
to the density contrast by:
\begin{equation}
	\label{xi_eq}
   \xi(r)=\frac{1}{3r^2}\frac{\mathrm{d}}{\mathrm{d}r}\left( r^3
   \Delta(r)\right) \, .
\end{equation}
In this framework, given a profile $\Delta(r)$ we can compute the
corresponding model for the $\xi(\sigma,\pi)$. 
However, the estimate of the correlation function using real data 
(see Section \ref{S_xi_data}) involves the use of a centre void sample. 
Thus, the profile $\Delta(r)$ should
be understood as the mean density profile of the void sample. 
In order to implement our model on observational data,
it is important to select samples of voids which share a similar profile,
well represented by the averaged one. 
In the following subsection, we define a parametric model for
this density profile.

\subsection{Velocity and density profile model}
\label{ss_voidflow}

In order to model the integrated density profiles of voids, we 
introduce a simple empirical model that contains all the necessary
features.
The R-type voids (as defined in Section \ref{S_xi_data}) have the
simplest profile shapes, a continuously rising
curve from zero to the mean density of the universe around the void radius.
The error function $\mathrm{erf}(x)$, behaves similarly, therefore we
choose this functional form to model such profiles,
\begin{equation}
	\label{eq_rho_r}
	\Delta_R(r) =
   \frac{1}{2}\left[\mathrm{erf}\left(\mathrm{S}\;\mathrm{log}(r/\mathrm{R})\right)-1\right].
\end{equation}
This model depends on two parameters, the void radius R and a
\textit{steepness} coefficient S.
On the other hand, the profiles of S-type voids are a bit more
complex and require two additional parameters in order to account
for the overdensity shell surrounding the void.
We add to the rising term in Eq. \ref{eq_rho_r}, an additional term representing
the peak on density due to this shell. 
Thus, the overdensity model for S-type void profiles is given by:
\begin{equation}
	\label{eq_rho_s}
	\Delta_S(r) =
   \frac{1}{2}\left[\mathrm{erf}\left(\mathrm{S}\;\mathrm{log}(r/\mathrm{R})\right)-1\right] + 
   \mathrm{P}\,\mathrm{exp}\left(-\frac{\mathrm{log^2}(r/\mathrm{R})}{2\Theta^2(r)}\right)
\end{equation}
where the Gaussian peak has an asymmetric width, 
\begin{equation}
   \Theta(r) = \left\{
   \begin{array}{ll}
      1/\sqrt{2\;\mathrm{S}} & r<\mathrm{R}\\
      1/\sqrt{2\;\mathrm{W}} & r>\mathrm{R}
   \end{array}
\right.\nonumber
	\label{eq_rho_Theta}
\end{equation}
As can be seen, such asymmetry is obtained by placing two semi-gaussians
instead of just one.
This allows us to modify the size of the shell, through the W parameter,
without changing the inner shape of the profile, related to S. 
Therefore, an S-type integrated overdensity profile requires the use of four 
parameters, namely R, S, P and W.
With this profile in equations \ref{v_eq} and \ref{xi_eq} it is possible to
compute the integral in Eq. \ref{xi_int}.
In order to perform this integration we use a Runge-Kutta method of sixth
order.
This allows a robust estimation of the integral avoiding numerical issues 
related with the rapid variation of the argument function. 
A schematic view of the role of the parameters on the model is shown 
in the upper right panel of Fig. \ref{lklyhd_12-10_12-over_mock}.
The S-type profile (solid line in the upper right panel of Fig. \ref{lklyhd_12-10_12-over_mock}) is
obtained from the expression \ref{eq_rho_s}. The rising term is shown as a 
long-dashed line, whereas the peak term is represented as short-dashed lines.

\subsection{Likelihood sampling and confidence intervals}
\label{ss_mcmc}

In the previous subsection we have presented the procedure to compute the 
model for the redshift space correlation function. 
In the current subsection we will describe the methods employed to determine
the best set of parameters which reproduce a given observed correlation
function on real or mock data.
To this end we have implemented a Markov Chain Monte Carlo method (hereafter
MCMC) to map the likelihood function of the $\xi$ model given the
corresponding measurements in a void sample.

The likelihood function compares the modelled correlation functions obtained
from different sets of parameters to the measured correlation functions of a given data
set, by quantifying the difference between them.
The MCMC method samples the posterior probability distribution of the
model given the data through a set of markov chains, which traverse
the parameter space until they reach the equilibrium distribution.
To explore this space, we employ the Metropolis-Hastings algorithm to
obtain a random sample of estimates of the model probability. 
In this process, the likelihood function allows to decide when a given
set of parameters is better at describing the observed correlations
than a previous set.
To properly quantify these model-data distances, we
estimate the covariance matrix of the observed correlation function.  This matrix
plays the role of a metric in the model parameter space.
The $\xi(\sigma,\pi)$ function is measured at $15\times15$ logarithmic
bins over the scale intervals used for each sample.
Thus, the
covariance matrix between each pair of bins in the correlation matrix,
hereafter denoted as $\mathbfss{C}$,
is a squared matrix of $15^2\times15^2$ elements.
Each element $\mathrm{C}_{ij}$ is the estimator of the variance,
computed on the data by jacknife resampling \citep{tukey_bias_1958}
using the multivariate generalization given by \citet{efron_jackknife_1987}:
\begin{equation}
	\mathrm{C}_{ij}=\frac{n-1}{n}\sum_{k=1}^n
	\left[\mathbf{\xi_{(k)}-\xi_{(.)}}\right]_i
	\left[\mathbf{\xi_{(k)}-\xi_{(.)}}\right]_j\,,
\end{equation}
where 
$n$ is the number of jackknife realizations,
$\mathbf{\xi_{(k)}}$ is the correlation function for the $k$th jackknife
realization and $\xi_{(.)}$ is the average
of $\mathbf{\xi_{(k)}}$ over the $n$ realizations.
The matrix $\mathbfss{C}$ is not diagonal, since the 
independence of the correlation values at bins in different scales
can not be guaranteed.
The probability $\mathscr{L}$ that a given model reproduces the data
results is then given by
\begin{equation} 
  \mathrm{Ln}(\mathscr{L}) = 
  -\mathbf{\Delta\xi}\cdot\mathbfss{C}^{-1}\mathbf{\Delta\xi} + \mathrm{const}\,,\nonumber 
\end{equation}
where $\mathbf{\Delta\xi}$ is a vector containing the differences
between the data and modelled correlation functions.

The computation of the likelihood depends on the inverse of the
covariance matrix.
However, instead of computing $\mathbfss{C}^{-1}$ it is more accurate to
solve the system $\mathbfss{C}\mathbf{a}=\mathbf{\Delta\xi}$ for the vector $\mathbf{a}$,
obtaining $\mathscr{L}$ as the inner product
$\mathbf{a}\cdot\mathbf{\Delta\xi}$.
Another numerical issue arises from the fact that the adopted estimator
for the covariance gives by definition a positive semidefinite matrix. 
The covariance matrix takes the form of a sparse matrix, due to fact
that the covariance of bin pairs at increasing separations approaches
zero, but fluctuates due to the noise introduced by the covariance
estimator.
This leads, in some cases, to a solution 
which can be dominated by numerical noise or may even
not exist.
This issues can be overcome by "tapering" the covariance matrix, i.e.
nullifying the covariance elements at large separations.
Following \citet{kaufman_covariance_2008}, we multiply element-wise
the covariance matrix estimated with a correlation matrix, defined to
force null values for elements with pair bin distances (calculated in
the two dimensions, parallel and perpendicular to the line of sight)
larger than 4
bins.
The tapering interval of 4 bins is large enough to leave unaltered
the principal features of the covariance matrix, ensuring at the same
time a positive-definite system.
We then compute the
probability for any given model in the parameter space using the 
tapered covariance matrix.
This procedure allows us to obtain an estimate of the parameters that
maximizes the Likelihood function and its corresponding confidence
intervals.
We use flat priors for all the parameters in the model, restricting
the search in the parameter space to a region where the model gives
meaningful profiles.

We show in Fig. \ref{lklyhd_12-10_12-over_mock} an example of the
fitting procedure used on a sample of S-type voids taken from the mock
catalogue.
These voids were identified over the M2 sample of the mock catalogue,
with radii ranging from 10 to \mbox{12 \dunit}.
The $\xi(\sigma,\pi)$ function was estimated following the methodology
described in section \ref{S_xi_data}.
In the upper right panel, we schematically represent the parameters
involved in the model of the integrated density void profiles
$\Delta(r)$, as it is described in the subsection \ref{ss_voidflow}.
We run 40 independent chains to explore this parameter space.
The convergence criterion is based on \citet{gelman_inference_1992},
which compares the spread in the means between chains to the variance
of the target distribution.
Once a given chain satisfies this criterion, 
we split it in two parts, discard the first half (ordered by step), 
and use the rest to map the likelihood function.
With this procedure we avoid the early stages of the random walk,
where the distribution of points in the parameter space does not
necessarily follow the equilibrium distribution.
In the diagonal panels of Fig. \ref{lklyhd_12-10_12-over_mock} we show
the one-dimensional marginalized constraints over each one of the four
parameters (R, S, P, W).
We also show the constraints on all pairs of parameters, indicating
the 68.3\%, 95.5\% and 99.7\% confidence intervals.
As can be seen the marginal distributions for each parameter resemble
Gaussian probability densities, thus the uncertainties for each
parameter are nearly symmetric. 
The two dimensional projections of the likelihood function exhibit a
well defined maximum, whereas its isoprobability contours indicate
that there are not significant degeneracies.

In the bottom panels of Fig. \ref{F_SDSS_xisigpi}, we show two
examples of correlation function models obtained from fits run over
SDSS results (upper panels).
These $\xi(\sigma,\pi)$ functions are obtained from the model with the
set of parameters for which the likelihood, with the corresponding
correlation measurements, reaches its maximum. 
As can be seen in a comparison between the upper and lower panels of
this figure, the proposed model seems to reproduce the more important
features observed in the measured correlations.
Moreover, in the lower left panel the white solid line  shows two
isocorrelation levels ($\xi \approx\,0.5,\,-0.2$) which clearly
manifest the expected dynamics for small voids.
For instance, the contour corresponding to $\xi\approx-0.2$ reaches
the abscissa axis at \mbox{$\sigma\approx$ 7 \dunit}, whereas along
the $\pi$ direction the contour seems to elongate up to around \mbox{8
\dunit}.
This can be thought as a distortion produced in the inner scales of
voids by outflow velocities of about \mbox{100 $\kms$ h$^{-1}$}. 
The opposite behaviour can be seen at the inner part of the isocontour
level of $\xi\approx 0.5$. This contour starts at
\mbox{$\sigma\approx$ 11 \dunit} and reaches the $\pi$ axis at less
than \mbox{9 \dunit}.
This can be interpreted as a contraction in the correlation function
contours due to the presence of redshift distortions, in this case
originated by infall velocities around \mbox{-100 $\kms$ h$^{-1}$}.
This kind of velocity profiles of outflowing velocities inside the
void region and infalling velocities at the outskirts, are expected in
smaller voids. This is in qualitative agreement with $\Lambda$CDM
predictions (see section \ref{S_intro}).
On the other hand for larger voids, as shown in the lower right panel
of Fig.  \ref{F_SDSS_xisigpi}, the model exhibits elongated contours
in the inner region. 
More precisely, the inner contour starts at $\sigma\approx 15$ \dunit
and reaches the $\pi$ axis around \mbox{16 \dunit}.
Although the profiles of R-type voids do not reach a maximum at any
scales, the modelled $\xi(\sigma,\pi)$ function for this sample
exhibits a clear maximum  located along the ordinate axis at
\mbox{$\pi\approx$ 27 \dunit}.
This maximum is surrounded by the $\xi\approx0.15$ contour which
encircles a postive correlation region breaking the isotropy of the
correlation map. 
As discussed in section \ref{S_xi_data} in the case of the observed
correlation function (upper right panel in Fig. \ref{F_SDSS_xisigpi})
such anisotropy could be thought as evidence of redshift space
distortions at the void outskirts. 

\begin{figure*}
\includegraphics[width=0.75\textwidth]{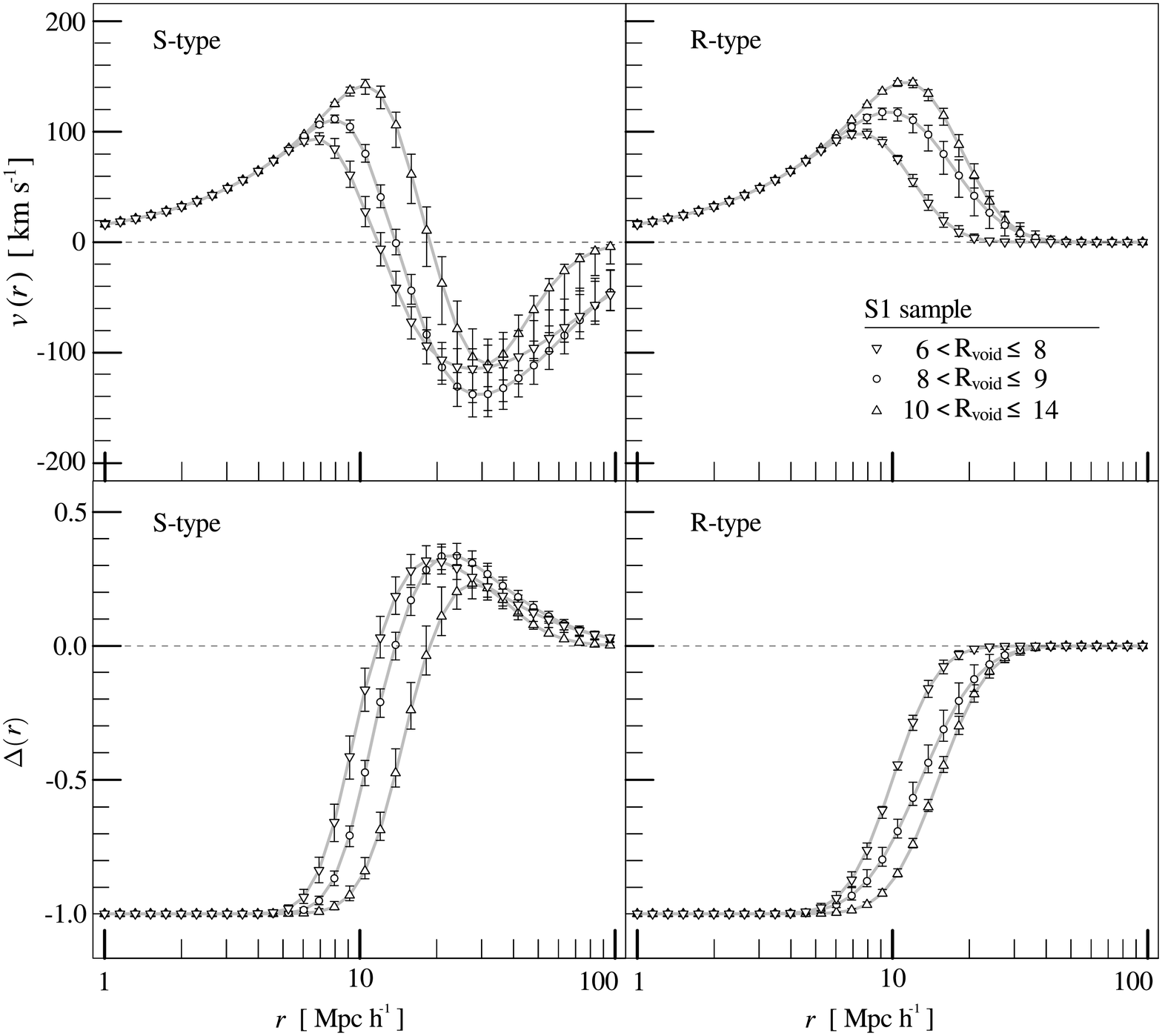}%
\caption{
Void--centric radial galaxy density profiles $\Delta(r)$ (lower
panels) and void--centric radial galaxy velocity profiles (upper panels) 
for S-type (left) and R-type (right) voids in the S1 sample. 
Different void radii ranges are indicated with downward triangles
(\mbox{6--8 \dunit}), circles (\mbox{8--9 \dunit}) and
upward triangles (\mbox{10--14 \dunit}).
Error bars indicate the region enclosing all curves within 68.3\%
uncertainty in parameter space.
}
\label{F_SDSS_Profile_Velocity}
\end{figure*}

In Section \ref{S_results} we provide an analysis of these results based on the
modelled velocity curves, in particular we show in Fig.
\ref{F_SDSS_Profile_Velocity} the corresponding profiles of these two void 
samples among others. 
The downward triangles in the left panels of this figure (S-type labeled) 
correspond to the small S-type voids in the left panels of Fig. 
\ref{F_SDSS_xisigpi}).
The upward triangles in the right panels of Fig. \ref{F_SDSS_Profile_Velocity} 
are the derived velocity (upper panel) and density curves (lower panel) from
the large R-type voids in the right panels of Fig. \ref{F_SDSS_xisigpi}). 
For further comments and a more detailed analysis please refer to section 
\ref{S_results}, where we also analyse the other SDSS void samples 
shown in the figure. 

In this section we have presented an analytic model for redshift
space distortions on the void-galaxy cross-correlation function.
We showed in this subsection how the parameters 
of this model can be obtained from fitting the redshift space correlation function. 
The likelihood and confidence intervals shown in Fig. \ref{lklyhd_12-10_12-over_mock} 
are a representative example of the results obtained for the 
different samples used in this work.
In the following section we provide an analysis of
how well this technique can be used to recover the real
velocity and density profile of voids from redshift-space data.

%}}}*/

% FIG 3
%{{{*/
\begin{figure*}
\includegraphics[width=0.75\textwidth]{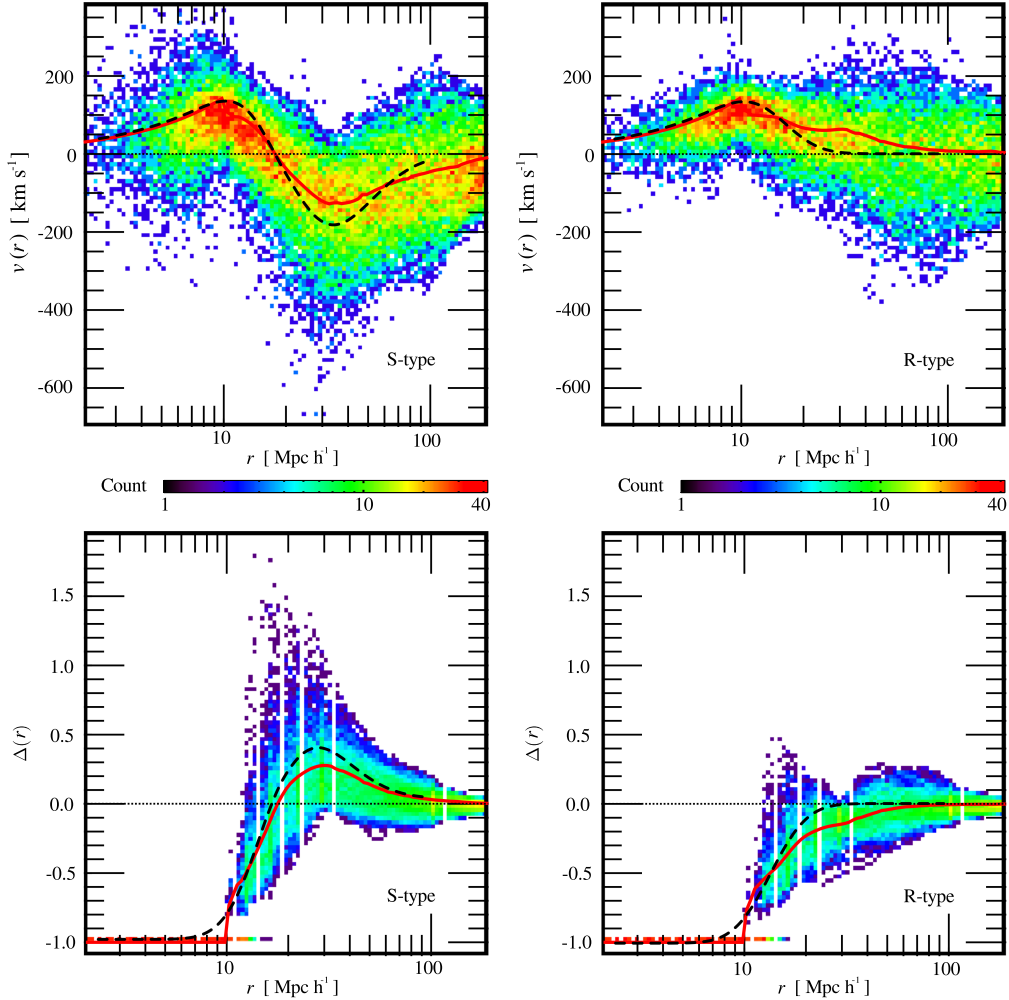}%
\caption{
   Two--dimensional histograms (color map) of distance-velocity (upper
   panels) and distance-density (lower panels) void profiles in the
   semi-analytic sample of galaxies of the simulation box.
   The color scale displays the number of profiles (integrated density
   and radial velocity, at lower and upper panels, respectively)
   overlapping at a given bin in radial distance and velocity or
   density.
   Centre voids have been identified in real space with radii in the
   range \mbox{10--12 \dunit} and separated into S (left) and
   R-type (right) samples.
   On each panel, the red solid line shows the averaged values on
   distance bins of velocity and density.
   For comparison, we also show the recovered velocity and density
   curves from the $\xi(\sigma,\pi)$ model in an equivalent mock
   sample (black dashed line).
   In this case, voids have been identified in the M2 sample of the
   mock catalogue with radii in the same range (\mbox{10--12
   \dunit}).
}
\label{F3} \end{figure*}
%}}}*/

\section{Testing the method on simulations}
\label{S_test}
%{{{*/

In the previous section we have presented a parametric model
for the $\xi(\sigma,\pi)$ function. 
We showed the corresponding model results for large 
and small voids in the SDSS, corresponding to 
R- and S-types respectively.
The model reproduces the main features of the observed
redshift space correlation functions for both samples.
Also, the model is characterized by a well behaved likelihood function, 
with not appreciable degeneracies in the parameter space. 
This suggests that our model  not only reproduces 
the observables but also gives a meaningful set of best-fitting parameters. 
Given that the model is physically motivated, its results can be used
to get insights about the dynamics of the large scale structure around voids. 
In this section we study the capacity of our model to recover the
underlying velocity and density profiles.
This analysis is performed by comparing the model results
obtained in the mock catalogue with direct measurements
of velocity and density profiles in the corresponding simulation. 
It will also allow us to quantify the effects of observational 
biases in the results.

In the Fig. \ref{F3} we show the results of the proposed test
for voids with radii between 10 and \mbox{16 \dunit}.
Centre voids have been identified in real space and separated into S (left panels) and R-type (right panels)
samples.
For each void in the simulation box we measured
its radial velocity, $v(r)$, and integrated density profiles,
$\Delta(r)$, in spherical shells.
We adopt negative values for inward radial velocities, whereas
positive values indicate outflowing velocities.
In the upper panels of the figure we display the number of radial
velocity curves overlapping a given bin in radial distance and radial
velocity respecting to the void centre.
The spread in radial velocity from void to void at fixed radial distance
can be seen from the color map, where redder colours indicate larger number
of curves (up to 40).
The red solid lines represent the mean velocity profiles for each sample,
which are close to the larger concentration of curves, indicating a nearly symmetric spread.
We also show, in the bottom panels, the results for the integrated radial density profiles.
Here again the color map indicate the number of curves, in this case integrated
density profiles, in radial distance and density bins. 
Red solid lines display the mean density curves for each sample (S and R-type voids at left and right panels, respectively).
Finally, we compare these simulation results with those estimated by using our model
in redshift space data.
We show, with dashed black lines, the velocity and density profile estimates obtained by applying our model in the
mock catalogue, following the procedure described in Section \ref{S_model}.
In synthesis, for S and R-type voids in the M2,
we compute the redshift space void-galaxy correlation function,
$\xi(\sigma,\pi)$, and its corresponding model fits.

As can be seen, for both velocity and density profiles the fitting procedure is
successful in recovering the underlying behaviour in the simulation. Some
differences can be seen at separations larger than the void radius and are due
to the limitation of the model for the density profile to reproduce the
detailed shape of the actual profile.  This of course translates into some
difficulty in reproducing the velocity profiles.
However, the overdensity values and the mean velocity differences are smaller
than $0.3$ and  \mbox{100 $\kms$} respectively, and the model successfully
recovers the S- or R-type nature of the profiles in both density and velocity.
In the following section we will apply this procedure to SDSS voids.

%}}}*/

% FIG 4
%{{{*/
%}}}*/

% Table 2
%{{{*/

\begin{table*}
\begin{minipage}{0.9\textwidth}
\begin{tabular*}{0.9\textwidth}{@{\extracolsep{\fill}}cccccccc}
\cline{1-8}
sample &  r$_{min}$ & r$_{max}$ & type &R & S & P & W \\
\cline{1-8} \\

  S1 &  6 & 8  & R & 10.06$\pm$ 0.06 & 5.1 $\pm$ 0.3 & - & - \\
     &    &    & S & 11.0$\pm$ 0.3 & 5.7 $\pm$ 0.3 & 0.42 $\pm$ 0.03& 1.77 $\pm$ 0.09\\
     &  8 & 9  & R & 12.8$\pm$ 0.3 & 3.9 $\pm$ 0.3 & - & - \\
     &    &    & S & 13.4$\pm$ 0.4 & 5.5 $\pm$ 0.3 & 0.46 $\pm$ 0.04& 2.0 $\pm$ 0.1\\
     & 10 & 14 & R & 15.0$\pm$ 0.2 & 4.6 $\pm$ 0.2 & - & - \\
    &    &    & S & 18.5$\pm$ 0.9 & 4.8 $\pm$ 0.4 & 0.58 $\pm$ 0.04& 2.7 $\pm$ 0.1\\
                   
  S2 & 9  & 12 & R & 13.5 $\pm$ 0.2 & 5.0 $\pm$ 0.2 & - & -\\
     &    &    & S & 15.1 $\pm$ 0.4 & 4.9 $\pm$ 0.2 & 0.39 $\pm$ 0.03& 1.83 $\pm$ 0.09\\
     & 12 & 15 & R & 16.5 $\pm$ 0.2 & 4.8 $\pm$ 0.1 & - & -\\
     &    &    & S & 17.7 $\pm$ 0.4 & 6.1 $\pm$ 0.3 & 0.32 $\pm$ 0.03 & 1.7 $\pm$ 0.1\\
     & 15 & 25 & R & 22.4 $\pm$ 0.3 & 4.0 $\pm$ 0.1 & - & -\\
                   
S3 & 11 & 14 & R & 14.5 $\pm$ 0.1 & 4.9 $\pm$ 0.1& - & -\\
   &    &    & S & 17.0 $\pm$ 0.2 & 5.4 $\pm$ 0.2& 0.43 $\pm$ 0.02& 2.05 $\pm$ 0.07\\
   & 14 & 19 & R & 17.6 $\pm$ 0.1 & 4.3 $\pm$ 0.1& - & -\\
   & 19 & 26 & R & 25.5 $\pm$ 0.1 & 4.01 $\pm$ 0.09& - & -\\

\\ \cline{1-8}
\end{tabular*}
\caption{
   Model parameters
   (see Eqs.  \ref{eq_rho_r}, \ref{eq_rho_s} and \ref{eq_rho_Theta})
   obtained for the different subsamples of SDSS voids.
   Minimum and maximum radii ($r_{min}$ and $r_{max}$, respectively) 
   are expressed in units of \dunit.
}
\label{T_results} 
\end{minipage}
\end{table*}

%}}}*/

\section{SDSS results}  \label{S_results}
%{{{*/

The large--scale region around voids determines two different populations 
of voids.
This was predicted from theoretical considerations in 
\citet{sheth_hierarchy_2004}, who also found that void environments
are a key factor in their dynamical behaviour.
In Paper I we found that small voids are likely to be surrounded by 
overdense shells, whereas larger voids tend to show smoothly rising
profiles.
According to the previously mentioned results, a different dynamical
behaviour of S and R-type voids is expected.
This has been studied for samples of voids derived 
from numerical simulations, both in the full simulation box and in
mock catalogues (including Paper I). 
However, the corresponding analysis in observational samples has not
yet been carried out, and this paper aims at confronting the
observations to the theoretical model expectations.
Thus, we have applied the methods described in Section \ref{S_xi_data} to 
our void catalogues in SDSS.

In the lower panels of Fig. \ref{F_SDSS_Profile_Velocity} we show
the resulting void--centric radial galaxy density profiles,
$\Delta(r)$, for S-type (left) and R-type (right) voids in sample S1.
The upper panels show the corresponding void--centric radial 
galaxy velocity profiles. 
As is indicated in the figure, the different symbols correspond to
different ranges in void sizes.
We indicate with downward triangles the void radii in the range
\mbox{6--8 \dunit}, with circles voids with radii in the range
\mbox{8--9 \dunit}; and with triangles voids with radii in the
range \mbox{10--14 \dunit}.
The error bars in Fig. \ref{F_SDSS_Profile_Velocity} represent the
68.3\% uncertainties resulting from the MCMC likelihood mapping.
As it can be seen in the figure the modelled profiles of S and R-type
voids are satisfactorily recovered and describe the typical behaviour
of the two types of voids.
Indeed, the observed density profiles are consistent with the
modelled profiles within uncertainties (not shown for the sake of
simplicity). 
Regarding the velocity profiles (upper panels of Fig. \ref{F_SDSS_Profile_Velocity}), it can be seen that the S-type voids 
show two different dynamical regimes.
While inner regions are in expansion, the large--scale void walls  are
collapsing.
This is in agreement with the void--in--cloud scenario introduced by 
\citet{sheth_hierarchy_2004} and the direct measurements in our 
numerical simulations presented in Paper I. 
On the other hand, the fitted velocity profiles of R-type voids never exhibit infall velocities
as can be seen in the bottom-right panel if this figure.
This behaviour fits well with the void-in-void scheme, which indicates that voids 
embedded in low density large--scale regions are likely to be
expanding.
These results provide the first observational evidence of the two
processes involved in void evolution.
We also find that the behaviour of these profiles are different
as the void size increases. 
Where voids surrounded by overdense large--scale shells are
under contraction, voids laking this outer overdensity are 
usually expanding.
In this scenario, voids embedded in overdense environments are
dominated by gravitational collapse rather than by expansion.
Consequently, it is likely that many of the small voids with a surrounding
overdense shell have sank inward by the present epoch.
Larger voids, on the other hand, are probably expanding in concordance
with the formation of the large structures that shape them.

We applied this procedure to the R and S-type subsamples in SDSS and
mock catalogues described in Table \ref{T_SampleDef}.
In the Table \ref{T_results} we show the resulting model parameter
fits, along with their uncertainties, derived from each subsample.
The radii ranges have been chosen taken into account the distribution 
of void radii, so that the sample is in each case divided into three 
subsamples with nearly the same number of voids each. 
As can be appreciated in this table, the parameter values 
support the scenario of a dichotomy in void evolution. 

%}}}*/             

\section{Summary and conclusions}  \label{S_concl}
%{{{*/

We have performed a statistical study of the void phenomenon focussing
on the dynamics of the surrounding regions of voids.
We used samples of voids identified following the procedure described
in \citet{padilla_spatial_2005}.
We constructed catalogues of voids in the SDSS-DR7, as well as in mock catalogues and 
in the parent simulation box to test the effects of observational
biases.

We analyze the dynamics of voids with and without a surrounding overdense shell in
the SDSS, dubbed R-type and S-type respectively, following
Paper I. 
We find that small voids, which are more frequently surrounded by overdense
shells (see Paper I), are likely to be in a collapse stage. 
On the other hand larger voids are in expansion, due to they have a large fraction of R-type
profiles (Paper I).
Using a model based on the linear theory of gravitational collapse, we
model the void-galaxy cross-correlation function in redshift space to
take advantage of the redshift-space distortions to obtain the 
dynamical properties of galaxies around voids. 
The analysis of the mock catalogues shows that the model successfully recovers the underlying velocity 
and density profiles of voids from redshift space samples.
When applying this procedure to SDSS data, we obtained evidence of a twofold population of
voids according to their dynamical properties as
suggested on previous observational studies 
(Paper I)
and as predicted by theoretical 
considerations by \citet{sheth_hierarchy_2004}.
According to this, some voids show a continuously rising profile 
fitting within the void-in-void scheme proposed by 
\citet{sheth_hierarchy_2004}.
Our redshift space-distortion studies indicate that this type of
voids are likely to be expanding.
Small voids, on the other hand, are tipically surrounded by an
overdense shell and their redshift space distortions indicate that 
they are more likely to be collapsing.

We test and interpret our results by comparing SDSS results to a semi-analytic mock 
galaxy catalogue extracted from the Millennium simulation. 
Both the mock catalog and the observational results are in very good
agreement, providing additional support to the viability of a $\Lambda$CDM
model to reproduce the large scale structure of the Universe as
defined by the void network and their dynamics.

%}}}*/

\section*{Acknowledgments}
%{{{*/
%
This work has been partially supported by Consejo de Investigaciones
Cient\'{\i}ficas y T\'ecnicas de la Rep\'ublica Argentina (CONICET)
and the Secretar\'{\i}a de Ciencia y T\'ecnica de la Universidad
Nacional de C\'ordoba (SeCyT).  NP acknowledges support from Fondecyt Regular
$1110328$ and BASAL CATA PFB-06.

We thank the anonymous referee for useful suggestions that
significantly increase the correctness and quality of this work.

Plots are made using R software and post-processed with Inkscape.
Algebraic computations were made using \textsc{LAPACK} routines.

Funding for the SDSS and SDSS-II has been provided by the Alfred P.
Sloan Foundation, the Participating Institutions, the National Science
Foundation, the U.S. Department of Energy, the National Aeronautics
and Space Administration, the Japanese Monbukagakusho, the Max Planck
Society, and the Higher Education Funding Council for England. The
SDSS Web Site is http://www.sdss.org/.
The SDSS is managed by the Astrophysical Research Consortium for the
Participating Institutions. The Participating Institutions are the
American Museum of Natural History, Astrophysical Institute Potsdam,
University of Basel, University of Cambridge, Case Western Reserve
University, University of Chicago, Drexel University, Fermilab, the
Institute for Advanced Study, the Japan Participation Group, Johns
Hopkins University, the Joint Institute for Nuclear Astrophysics, the
Kavli Institute for Particle Astrophysics and Cosmology, the Korean
Scientist Group, the Chinese Academy of Sciences (LAMOST), Los Alamos
National Laboratory, the Max-Planck-Institute for Astronomy (MPIA),
the Max-Planck-Institute for Astrophysics (MPA), New Mexico State
University, Ohio State University, University of Pittsburgh,
University of Portsmouth, Princeton University, the United States
Naval Observatory, and the University of Washington.

The Millennium Simulation databases used in this paper and the web
application providing online access to them were constructed as part
of the activities of the German Astrophysical Virtual Observatory.
%}}}*/

%%% Bibliography /// /// /// ///
%%% \bibliographystyle{mn2e}
%%% \bibliography{draft2.bib}

\end{document}